\documentclass[prd,aps,twocolumn,nofootinbib,preprintnumbers,superscriptaddress,preprintnumbers,balancelastpage,longbibliography]{revtex4-2}

\usepackage{graphicx}
\usepackage{bm}
\usepackage{hyperref}
\usepackage{hyperref}
\usepackage{slashed}
\usepackage{comment}
\usepackage{aas_macros}
\usepackage{float}
\usepackage{slashed}
\usepackage{lipsum}
\usepackage{subfigure}
\usepackage{multirow}
\usepackage{amsmath}
\usepackage{array} 
\usepackage{varwidth} 
\usepackage{tabularx}
\usepackage{braket}
\usepackage[dvipsnames]{xcolor}

\usepackage[normalem]{ulem}

\newcommand{\be}{\begin{equation}}
\newcommand{\ee}{\end{equation}}
\newcommand{\bea}{\begin{eqnarray}}
\newcommand{\eea}{\end{eqnarray}}

\hypersetup{
     colorlinks   = true,
     citecolor    = Fuchsia,
     urlcolor     = Fuchsia,
     linkcolor    = Fuchsia
}

\usepackage{listings}
\usepackage{color,xcolor}

\begin{document}
\preprint{KCL-2025-30}


\title{INTEGRAL, eROSITA and Voyager Constraints on Light Bosonic Dark Matter: \\ ALPs, Dark Photons, Scalars, $B-L$ and $L_{i}-L_{j}$ Vectors}

\author{Thong T.Q. Nguyen}
\thanks{{\scriptsize Email}: \href{mailto:thong.nguyen@fysik.su.se}{thong.nguyen@fysik.su.se}; \href{https://orcid.org/0000-0002-8460-0219}{0000-0002-8460-0219}}
\affiliation{Stockholm University and The Oskar Klein Centre for Cosmoparticle Physics, Alba Nova, 10691 Stockholm, Sweden}

\author{Pedro De la Torre Luque}
\thanks{{\scriptsize Email}: \href{mailto:pedro.delatorre@uam.es}{pedro.delatorre@uam.es}; \href{https://orcid.org/0000-0002-4150-2539}{0000-0002-4150-2539}}
\affiliation{Stockholm University and The Oskar Klein Centre for Cosmoparticle Physics, Alba Nova, 10691 Stockholm, Sweden}
\affiliation{Departamento de F\'{\i}sica Te\'orica, M-15, Universidad Aut\'onoma de Madrid, E-28049 Madrid, Spain}
\affiliation{Instituto de F\'{\i}sica Te\'orica UAM-CSIC, Universidad Aut\'onoma de Madrid,\\
C/ Nicol\'as Cabrera, 13-15, 28049 Madrid, Spain}

\author{\mbox{Isabelle John}}
\thanks{{\scriptsize Email}: \href{mailto:isabelle.john@unito.it}{isabelle.john@unito.it}; \href{https://orcid.org/0000-0003-2550-7038}{0000-0003-2550-7038}}
\affiliation{Dipartimento di Fisica, Universit\`a degli Studi di Torino, via P.\ Giuria, 1 10125 Torino, Italy}
\affiliation{INFN -- Istituto Nazionale di Fisica Nucleare, Sezione di Torino, via P.\ Giuria 1, 10125 Torino, Italy}

\author{Shyam Balaji}
\thanks{{\scriptsize Email}: \href{mailto:shyam.balaji@kcl.ac.uk}{shyam.balaji@kcl.ac.uk};  \href{https://orcid.org/0000-0002-5364-2109}{0000-0002-5364-2109}}
\affiliation{Physics Department, King’s College London, Strand, London, WC2R 2LS, United Kingdom}

\author{Pierluca Carenza}
\thanks{{\scriptsize Email}: \href{mailto:pierluca.carenza@fysik.su.se}{pierluca.carenza@fysik.su.se};  \href{https://orcid.org/0000-0002-8410-0345}{0000-0002-8410-0345}}
\affiliation{Stockholm University and The Oskar Klein Centre for Cosmoparticle Physics, Alba Nova, 10691 Stockholm, Sweden}

\author{Tim Linden}
\thanks{{\scriptsize Email}: \href{mailto:linden@fysik.su.se}{linden@fysik.su.se};  \href{http://orcid.org/0000-0001-9888-0971}{0000-0001-9888-0971}}
\affiliation{Stockholm University and The Oskar Klein Centre for Cosmoparticle Physics, Alba Nova, 10691 Stockholm, Sweden}
\affiliation{Erlangen Centre for Astroparticle Physics (ECAP), FAU Erlangen-Nürnberg, Nikolaus-Fiebiger-Str. 2,
91058 Erlangen, Germany}

\begin{abstract}
The decay of light bosonic dark matter particles can produce a bright electron/positron ($e^+e^-$) flux that can be strongly constrained by local Voyager observations of the direct $e^+e^-$ flux, as well as 511~keV Line and X-ray continuum observations of $e^+e^-$ emission. We carefully analyze the $e^+e^-$ yield and resulting cosmic-ray and X-ray spectra from theoretically well-motivated light dark matter models, including: (a) electrophilic axion-like particles, (b) dark photons, (c) scalars, and (d) $B-L$ and $L_{i}-L_{j}$ vector bosons. We use the morphology and spectrum of the INTEGRAL 511~keV line data, the eROSITA X-ray continuum spectrum and the Voyager $e^+e^-$ spectrum to constrain the decay lifetime and coupling of each dark matter model. We find that 511~keV observations typically set world-leading limits on bosonic dark matter decay below masses of $\sim$1~GeV, while eROSITA observations provide the strongest constraints in the range from 1--10~GeV. Finally, we forecast future limits from 21~cm line searches with next-generation HERA data. 

\end{abstract}

\maketitle

\section{Introduction}
\label{sect:intro}

The search for non-gravitational interactions between dark matter and Standard Model (SM) particles is one of the most profound open questions in physics~\cite{Bertone:2016nfn}. One prominent strategy is to look for an indirect signal from dark matter annihilation or decay in astronomical environments. Dark matter decays can produce observable SM particles, including photons, antiprotons, positrons, or neutrinos that can be observed by both space-based and ground-based telescopes~\cite{Bertone:2018krk}. The combined efforts of astronomers and particle physicists have derived stringent constraints on the dark matter decay lifetime for dark matter masses spanning from below an eV up to the Planck scale~\cite{DelaTorreLuque:2025zjt, Roy:2023omw, Cirelli:2023tnx, Cirelli:2020bpc, Liu:2020wqz, Qin:2023kkk,Carenza:2023qxh, Foster:2021ngm, Jin:2013nta, Blanco:2018esa, Das:2024bed, Murase:2012xs, Song:2024vdc, Song:2023xdk, Bauer:2020jay, Cohen:2016uyg, Baldes:2020hwx, Cheung:2018vww}.

For dark matter masses below a few GeV, kinematic considerations significantly constrain the Standard Model final states from dark matter annihilation or decay. In this paper we study four well-motivated bosonic dark matter models in this mass range: electrophilic axion-like-particles~\cite{Han:2020dwo,Calibbi:2016hwq}, dark photons~\cite{Fabbrichesi:2020wbt, Hebecker:2023qwl, Fayet:1980rr, Fayet:1990wx}, scalar dark matter with Yukawa-like couplings~\cite{Nguyen:2024kwy}, and flavor-dependent vector dark matter particles (the $B-L$ and $L_{i}-L_{j}$ families)~\cite{Chun:2022qcg}. In this paper we focus on $e^+e^-$ production and calculate the full $e^+e^-$ spectra for each candidate, taking into account branching ratios for all relevant dark matter decay channels.

We constrain these models using three leading indirect probes of $e^+e^-$ production:

\begin{enumerate}
    \item[$\circ$] \emph{Voyager --- } The decay of light dark matter to $e^+e^-$ pairs can be directly probed by Voyager constraints on the local $e^+e^-$ flux near the Sun. Despite its age and small detection area, Voyager's position outside of the heliosphere makes it the leading probe of low-energy $e^+e^-$, due to the significant modulation of low-energy cosmic rays by solar winds~\cite{stone2013voyager, cummings2016galactic}. 
    
    \item[$\circ$] \emph{INTEGRAL --- } Low-energy positrons produced via dark matter annihilation can thermalize and form positronium with Galactic electrons, which subsequently decays into two-photon final states at a characteristic energy of 511~keV~\cite{Ascasibar:2005rw,Boehm:2002yz,Boehm:2003bt,Vincent:2012an,Wilkinson:2016gsy, Khan:2024biq, Aghaie:2025dgl, Calore:2021lih}. While observations of the 511-keV excess have previously been studied as potential signals of light dark matter~\cite{Boehm:2003bt, Cappiello:2023qwl}, recent studies have shown that the low observed flux at high galactic longitude conversely places very strong constraints on dark matter decay~\cite{DelaTorreLuque:2023cef, DelaTorreLuque:2024wfz}.

    \item[$\circ$] \emph{eROSITA --- } X-ray observations are also interesting to probe the injection of electrons in the Galaxy~\cite{Cirelli:2009vg, Cirelli:2020bpc, Cirelli:2023tnx, DelaTorreLuque:2023olp, DelaTorreLuque:2024qms}. They can upscatter the low-energy photon fields in the Galaxy and generate an X-ray emission via Inverse-Compton scattering (ICS). We make use of the recent eROSITA~\cite{2012arXiv1209.3114M, 2024A&A...681A..77Z, eROSITA:2020emt} diffuse data in the band between $0.2$ to $1$~keV.
\end{enumerate}

This paper is organized as follows. In Section~\ref{sect:models}, we review the dark matter models that we consider. We provide the decay branching ratios as functions of dark matter mass and coupling for all SM final states. In Section~\ref{sect:511keV}, we model $e^+e^-$ pair production from each model and calculate the strength and morphology of the resulting Voyager, INTEGRAL and eROSITA signals. In Section~\ref{sect:result}, we compare this observational data with the X-ray and cosmic-ray signals from each dark matter candidate, constraining both the lifetimes and the fundamental SM couplings of each dark matter model as a function of the dark matter mass.  We discuss the most important trends regarding the strength of these limits in Section~\ref{sect:discussion}. Finally, we summarize our study in Section~\ref{sect:conclusion} and discuss potential improvements in our sensitivity to dark matter models in the MeV range.

\section{Bosonic Dark Matter Models}
\label{sect:models}

We consider four classes of dark matter models: electrophilic ALPs, dark photons, flavour-dependent vector bosons, and scalar dark matter with Yukawa-like couplings. In the following sub-sections, we review the Lagrangian of each model and calculate their decay branching ratios to all SM final states.

\subsection{Electrophilic Axion-Like Particles}
\label{subsect:ALP}

Axion-like particles (ALPs) are hypothetical pseudo-Nambu-Goldstone bosons which emerge naturally in many SM extensions that involve spontaneous symmetry breaking in hidden sectors~\cite{Antel:2023hkf}. Besides the QCD axion, which was introduced to solve the strong-CP problem, ALPs with more general properties arise in various theories and might constitute viable dark matter candidates~\cite{DiLuzio:2020wdo,Jaeckel:2010ni,Svrcek:2006yi,Cicoli:2012sz,Halverson:2019kna,Carenza:2024avj,Carenza:2024qaq, Roy:2025mqw}.

The ALP-electron coupling makes it possible to design laboratory experiments to probe these particles~\cite{PandaX:2017ock,Majorana:2016hop,EDELWEISS:2018tde,SuperCDMS:2019jxx,XENON:2020rca,GERDA:2020emj,LUX:2017glr,PandaX:2024kjp,PandaX:2024cic,LZ:2023poo,DarkSide:2022knj,XENON:2022ltv} (see also Ref.~\cite{Ferreira:2022egk}). Even though we are primarily interested in ALPs that constitute dark matter, it is worth mentioning that MeV-scale electrophilic ALPs are effectively probed through observations of stellar properties~\cite{Carenza:2021osu,Carenza:2024ehj, Ghosh:2020vti,Carenza:2021pcm,Ferreira:2022xlw, Ning:2025tit, Fiorillo:2025sln}, blazar jets~\cite{Ghosh:2025rjh},  Big-Bang Nucleosynthesis and  the Cosmic Microwave Background (CMB)~\cite{Ghosh:2020vti,Depta:2020zbh} regardless of whether they are a viable dark matter candidate.

One prominent theoretical motivation for heavy leptophilic ALPs occurs in models where the ALP emerges from a broken symmetry in the lepton sector, potentially explaining the mass hierarchy of the charged leptons~\cite{Han:2020dwo,Calibbi:2016hwq}.
In such scenarios, the ALP couples to electrons via an interaction term of the form:  
\begin{equation}
\mathcal{L}_{a} \supset g_{ae}  \bar{e} \gamma^5 \gamma^\mu e \, \partial_\mu a,
\label{eq:lagrange_ALP}
\end{equation}
where $g_{ae}$ is the ALP-electron coupling, $a$ is the ALP field and $e$ denotes the electron field. 
The decay rate of an axion into an electron-positron pair is given by~\cite{DiLuzio:2020wdo}
\begin{align}
\label{eq:Width_ALP}
\Gamma_{a \to {\rm e}^+ {\rm e}^-}& = \frac{g_{ae}^2 m_a}{8\pi} \sqrt{1 - \frac{4m_{\rm e}^2}{m_a^2}}\\
&\simeq 6.1 \times 10^{-19}{\rm s}^{-1}\left(\frac{g_{ae}}{10^{-19}}\right)^2 \left(\frac{m_{a}}{{\rm MeV}}\right)\sqrt{1 - \frac{4m_{\rm e}^2}{m_a^2}},\nonumber
\end{align}  
where ALP decay is kinematically allowed for \mbox{$m_a>2m_{\rm e}$.} Enforcing stability on cosmological scales requires $\Gamma_{a \to {\rm e}^+ {\rm e}^-}\lesssim2.3\times10^{-18}~{\rm s}^{-1}$. Notably, electrophilic (and photophobic) ALPs can still decay into final states involving photons and neutrinos through loop-level processes. However, these channels are subdominant above the electron-positron pair production threshold~\cite{Bauer:2017ris}. Thus, $a\to{\rm e}^{+}{\rm e}^{-}$ is the dominant decay channel in our analysis.

\subsection{Dark Photon}
\label{subsect:darkphoton}

The dark photon is a hypothetical spin-1 particle that emerges from the inclusion of an extra $U(1)$-symmetry group~\cite{Fabbrichesi:2020wbt, Hebecker:2023qwl, Fayet:1980rr, Fayet:1990wx}. In the context of dark matter, scenarios such as inflation~\cite{Graham:2015rva}, extra-dimensions~\cite{Servant:2002aq}, misalignment with Steuckelberg mechanisms~\cite{Nelson:2011sf}, or string compactifications~\cite{Hebecker:2023qwl, Goodsell:2009xc} are proposed to explain the dark photon dark matter abundance in the universe. 

\begin{figure}[t]
    \centering
    \includegraphics[width=1\linewidth]{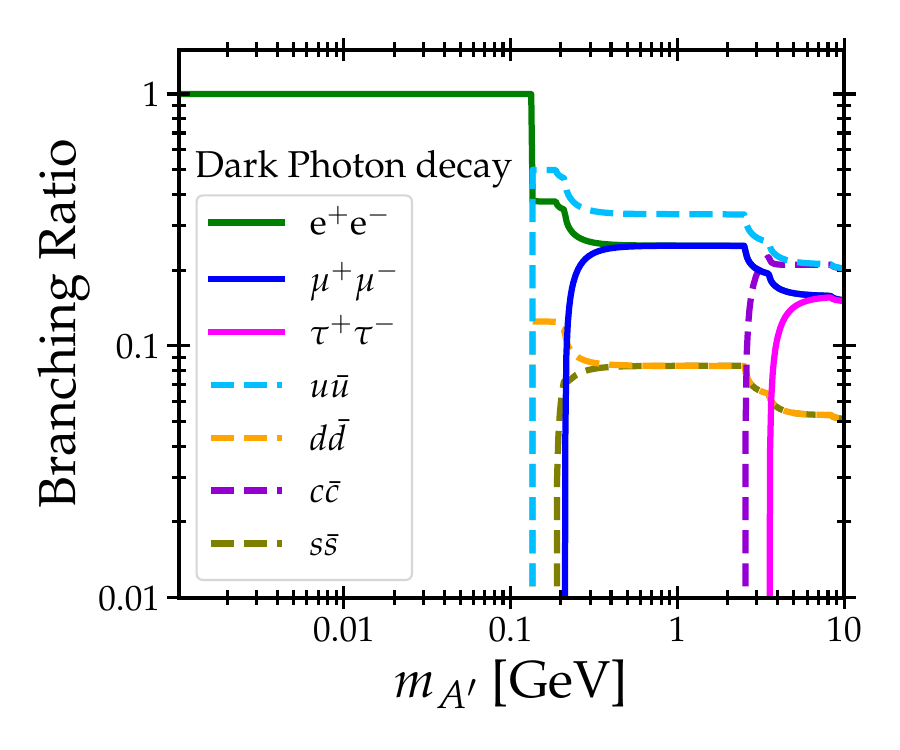}
    \vspace{-0.5cm}
    \caption{Branching ratios for all Standard Model decay final states of the dark photon particle. Solid lines are for charged leptons: electron (green), muon (blue) and tau (magenta). The dashed lines are for quarks: up (dodgerblue), down (orange), charm (purple) and strange (olive).} 
    \label{fig:BR_DP}
\end{figure}

Dark photons can interact with other SM particles through mixing with SM neutral bosons. For dark photon masses, $m_{A^{\prime}}$, that lie below the weak scale, the mixing term in the Lagrangian is
\begin{equation}
    \mathcal{L}_{A^{\prime}} \supset -\frac{\epsilon}{2}F^{\prime}_{\mu\nu}F^{\mu\nu}
    \label{eq:lagrange_DP}
\end{equation}
where $\epsilon$ is the kinetic mixing coupling, and $F^{\prime}_{\mu\nu}$ is the new dark photon field strength tensor, similar to the QED photon~\cite{Fabbrichesi:2020wbt}. The kinetic mixing coupling and the dark photon mass are targets for many experimental and phenomenological studies~\cite{Caputo:2021eaa} across astrophysical and laboratory searches~\cite{Redondo:2008ec, Arias:2012az, Bloch:2016sjj, XENON:2019gfn, XENON:2020rca, XENON:2021qze, An:2020bxd, Fischbach:1994ir, Zechlin:2008tj, Bi:2020ths, Tran:2023lzv, Li:2023vpv, Wadekar:2019mpc, Dolan:2023cjs, Dubovsky:2015cca, Yan:2023kdg, Hong:2020bxo, Vinyoles:2015aba, Carenza:2025uwx, Cyncynates:2024yxm, Kitajima:2024jfl, Linden:2024uph, Melo:2025prc, Fan:2024mhm, Fan:2022uwu, Jaeckel:2023huy, Nguyen:2025eva, Yin:2024lla,Balaji:2025alr,Carenza:2025uwx}.

Dark photons can decay to other SM particles if kinematically allowed. The decay rate for a specific fermion pair with mass $m_{f}$ in the final state is

\begin{equation}
    \Gamma_{A^{\prime}\to f\bar{f}}=\frac{N_{c}Q_{f}^{2}\alpha\epsilon^{2}}{3}m_{A^{\prime}}\sqrt{1-\frac{4m_{f}^{2}}{m_{A^{\prime}}^{2}}}\left(1+\frac{2m_{f}^{2}}{m_{A^{\prime}}^{2}}\right),
    \label{eq:Width_DP}
\end{equation}
where $N_{c}$ is the color number ($N_{c}=1$ for leptons, $N_{c}=3$ for quarks), and $Q_{f}$ is the fermion electric charge. Notably, despite the fact that dark matter with masses from 7--135~MeV are kinematically able to decay to up- and down-quark states, these particles lie below the QCD scale and are unable to produce the lightest bound states of these quarks, such as the neutral pion. Thus, for dark photons below 135~MeV, the electron-positron pair is the dominant decay channel.

We calculate the full branching ratios for dark photon decay to all SM final states. We show the results in Figure~\ref{fig:BR_DP}. Notably, there are two other final states from dark photon decay. Dark photons can decay into three photons through a loop-induced process, which is called the dark photon trident~\cite{Linden:2024fby,Cheung:2025gdn, Nguyen:2025eva}. This loop suppressed process is only relevant below 1~MeV. The dark photon can also decay to neutrinos through mixing with the SM $Z-$boson. However, this process is only relevant for heavy dark matter around the $Z$-mass~\cite{Nguyen:2024kwy}, while the decay width is suppressed in the mass range we consider (below 10~GeV).

\subsection{Scalar dark matter with Yukawa-like couplings}
\label{subsect:Scalar}

\begin{figure}[t]
    \centering
    \includegraphics[width=1\linewidth]{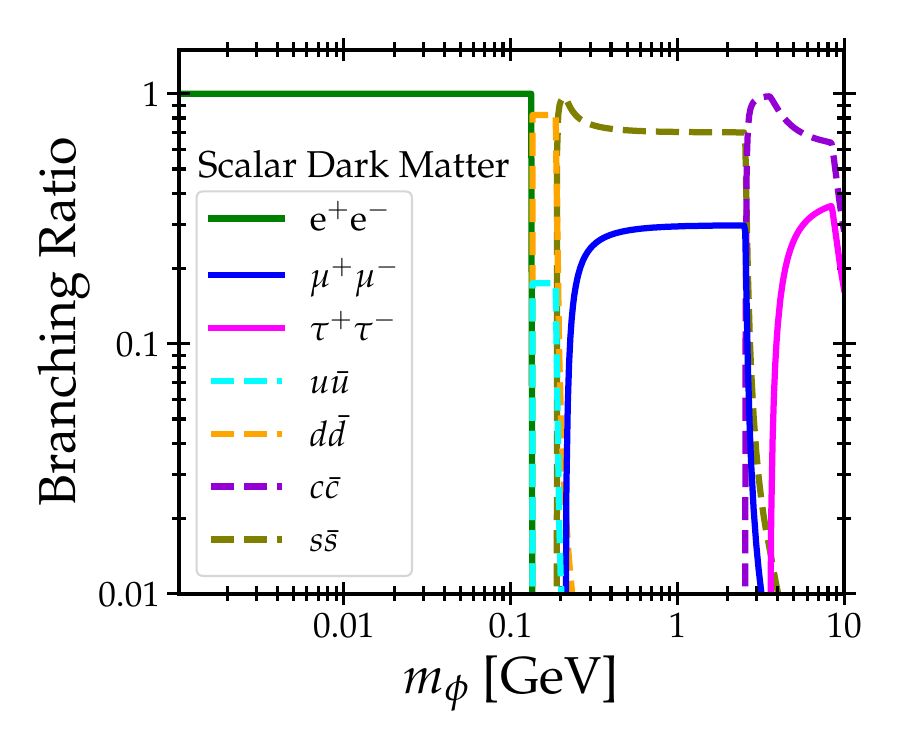}
    \vspace{-0.5cm}
    \caption{Branching ratio for all the SM decay modes of a scalar particle with Yukawa-like couplings, similar to Figures~\ref{fig:BR_DP} and \ref{fig:BR_Flavor}.} 
    \label{fig:BR_Scalar}
\end{figure}

\begin{figure*}[t]
\centering
\includegraphics[width=2\columnwidth]{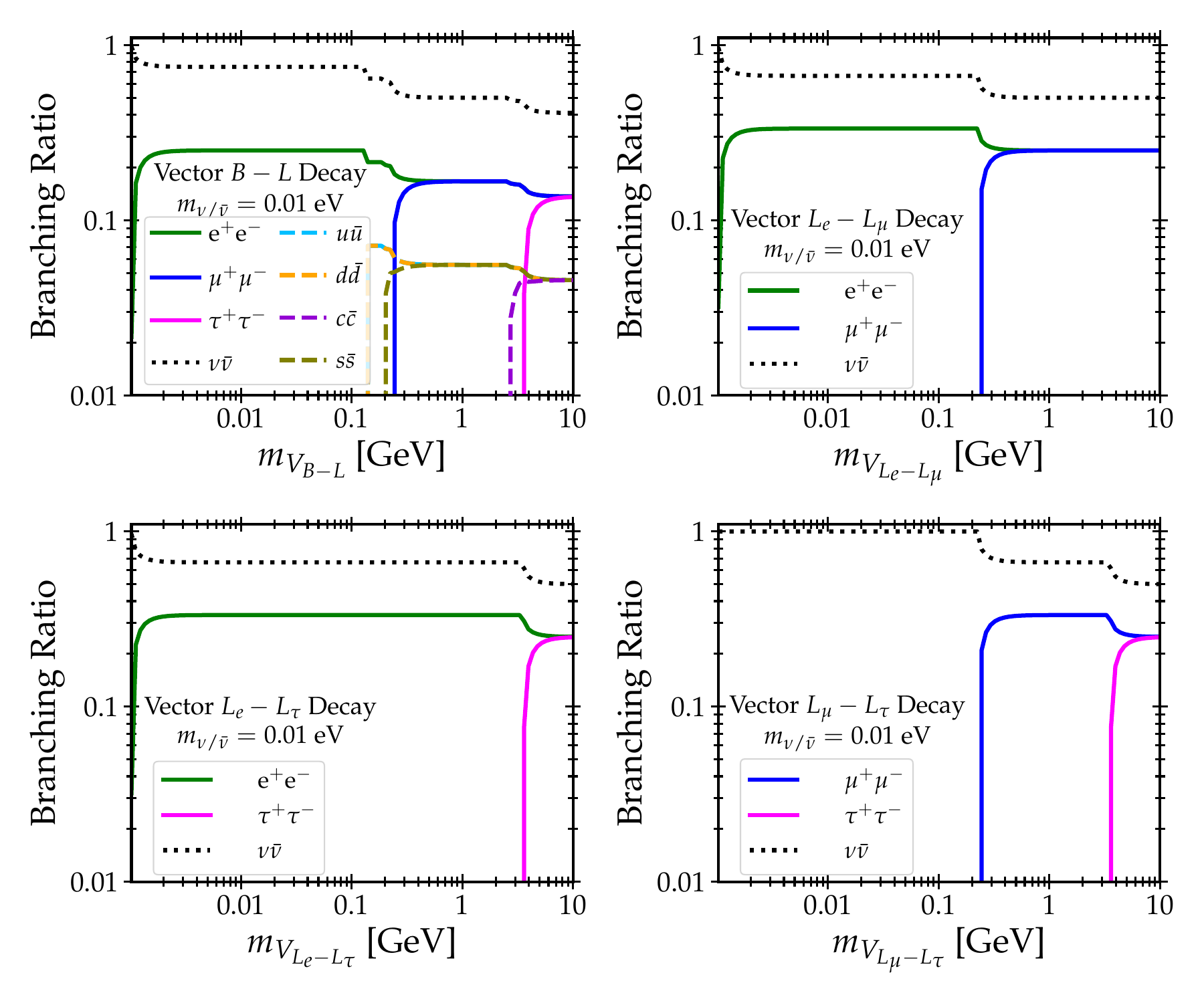}
\caption{Branching ratios of vector particle decays to the SM, with flavor-conserving interactions, similar to Figure~\ref{fig:BR_DP}: $B-L$ (top left), $L_{e}-L_{\mu}$ (top right), $L_{e}-L_{\tau}$ (bottom left) and $L_{\mu}-L_{\tau}$ (bottom right). The dominant branching ratio to neutrinos (which typifies these models) is shown as a black dotted line.  We choose all (anti-)neutrino masses to be $m_{\nu/\bar{\nu}}=0.01$~eV~\cite{DESI:2025ejh}. }
\label{fig:BR_Flavor}
\end{figure*}

The simplified model of a scalar dark matter candidate is discussed in Refs.~\cite{Boehm:2003hm, Kraml:2019sis, Dev:2020jkh, Bertrand:2020lyb,Bottaro:2023gep,Fiorillo:2025zzx,Wadekar:2021qae}.  The interaction between the scalar, $\phi$, and other SM fermions mimics the Yukawa interaction of the Higgs boson. The scalar mixes with the SM Higgs when the Higgs gets a vacuum expectation value (VEV). The mixing term is $C_{\phi}\phi|H|^{2}$, where $C_{\phi}$ is the coupling strength. Thus, the new scalar particle has a mass eigenstate that depends on the Higgs VEV and the scaled down Yukawa coupling. When $C_{\phi}$ is small enough, the Yukawa-like couplings of the scalar are smaller than the Higgs couplings, ensuring that the $\phi$-boson does not affect other SM particle masses.

The Lagrangian that describes this new scalar interaction with SM fermions is given by
\begin{equation}
    \mathcal{L}\supset -\sin \theta \frac{m_{f}}{v}\phi\bar{f}f,
    \label{eq:lagrange_scalar}
\end{equation}
where the Higgs VEV is $v\approx 246$~GeV, and $\theta$ is the mixing angle between $\phi$ and the SM Higgs. This interaction is proportional to the SM mass $m_{f}$. The decay width of $\phi$ to SM fermions is

\begin{equation}
    \Gamma(\phi\to \bar{f}f)=\frac{N_{c}G_{F}m_{f}^{2}}{4\sqrt{2}\pi }m_{\phi}\sin^{2}\theta\left(1-\frac{4m_{f}^{2}}{m_{\phi}^{2}}\right)^{3/2},
    \label{eq:Width_Scalar}
\end{equation}
where $G_{F}\simeq 1.66\times 10^{-5}$~GeV$^{-2}$ is the Fermi constant. Notably, $\phi$ can also decay to $W$ and $Z$ bosons. However, since we consider dark matter masses below 10~GeV, these decay channels are kinematically forbidden.

In Figure~\ref{fig:BR_Scalar}, we show the full branching ratio for all SM decay channels of $\phi$ above twice the electron mass\footnote{In our analysis, $\phi$ can also decay to massive neutrinos. However, the couplings are so suppressed that we do not take them into account, and assume the branching ratio into the $e^+e^-$ channel below the QCD scale to be 100\%, similar to previous studies.}. We apply these branching ratios in the context of $\phi$ decay. Notably, the literature has considered this dark matter particle in the light to ultralight mass regime, where the dominant interaction is with electrons~\cite{Branca:2016rez, Filzinger:2023qqh, Tretiak:2022ndx, Savalle:2020vgz, Vermeulen:2021epa, Aiello:2021wlp, Campbell:2020fvq, Oswald:2021vtc, Kennedy:2020bac, Zhang:2022ewz, Gottel:2024cfj, NANOGrav:2023hvm, Zhou:2025wax, Cyncynates:2024bxw, Cyncynates:2024ufu}. Our analysis focuses on heavier scalars that can both decay into $e^+e^-$ and can also interact with other SM particles.

\subsection{Vector dark matter: $B-L$ and $L_{i}-L_{j}$}
\label{subsect:flavor}

Flavor conserved $B-L$ and $L_{i}-L_{j}$ vector particles are also consequences of $U(1)$ extensions of the SM. Their quantum numbers depend on the baryon and lepton numbers of the SM particles they interact with. Similar to the dark photon, these particles are dark matter candidates in many production scenarios~\cite{Graham:2015rva, Servant:2002aq, Nelson:2011sf, Hebecker:2023qwl}. In terms of beyond the SM model building, these families of $U(1)$ flavor conserved gauge symmetries are anomaly-free with respect to the SM and there is no need to add extra fermions for UV completion. Hence, these models are considered to be well-motivated and have been probed in numerous experiments~\cite{Sushkov:2011md, Wagner:2012ui, Adelberger:2009zz, Chun:2022qcg, Nguyen:2022zwb, Li:2023vpv, Asai:2022zxw, Hayashi:2024not}.

For the $B-L$ gauge symmetry group, the Lagrangian describing the interaction between the $B-L$ vector particle and SM fermions is given by:
\begin{equation}
    \mathcal{L}_{B-L}\supset g_{B-L}V_{B-L}^{\mu}\sum\limits_{f}Q_{B-L}^{f}\bar{f}\gamma_{\mu}f,
    \label{eq:lagrange_BL}
\end{equation}
\noindent for a coupling $g_{B-L}$ and baryon-minus-lepton number charge $Q_{B-L}^{f}\equiv B_{f}-L_{f}$. We note that the baryon number for quarks is $B_{q}=1/3$ and the lepton number of any lepton is $L_{\ell,\nu}=1$. For a massive vector $B-L$ particle with mass $m_{B-L}$, the decay width to any SM fermion pair is:

\begin{equation}
    \begin{split}
    \Gamma_{V\to f\bar{f}}^{B-L}=&\frac{N_{c}g_{B-L}^{2}(Q^{f}_{B-L})^{2}}{12\pi}m_{V_{B-L}}\\
    &\times\left(1+\frac{2m_{f}^{2}}{m_{V_{B-L}}^{2}}\right)\sqrt{1-\frac{4m_{f}^{2}}{m_{V_{B-L}}^{2}}}.
    \end{split}
    \label{eq:Width_BL}
\end{equation}

Similarly, for the $L_{i}-L_{j}$  particle that only interacts with SM leptons, the interaction Lagrangian is:

\begin{equation}
\mathcal{L}_{L_{i}-L_{j}}\supset g_{L_{i}-L_{j}}V_{L_{i}-L_{j}}^{\mu} \sum\limits_{\ell=\ell_{i},\ell_{j}}Q^{\ell}_{L_{i}-L_{j}}\bar{\ell}\gamma_{\mu}\ell,
\label{eq:lagrange_LiLj}
\end{equation}
with $\ell$ denoting the lepton field. The decay width to an arbitrary leptonic final state is 

\begin{equation}
    \begin{split}
    \Gamma_{V\to f\bar{f}}^{L_{i}-L_{j}}=&\frac{g_{L_{i}-L_{j}}^{2}(Q^{f}_{L_{i}-L_{j}})^{2}}{12\pi}m_{V_{L_{i}-L_{j}}}\\
    &\times\left(1+\frac{2m_{f}^{2}}{m_{V_{L_{i}-L_{j}}}^{2}}\right)
    \sqrt{1-\frac{4m_{f}^{2}}{m_{V_{L_{i}-L_{j}}}^{2}}}.
    \end{split}
    \label{eq:Width_LL}
\end{equation}
\noindent where $g_{L_{i}-L_{j}}$ and $m_{V_{L_{i}-L_{j}}}$ represent the coupling and mass of the new gauge boson. The indices $i$ and $j$ ($i\neq j$) run from 1 to 3, indicating the interactions with $e^{\pm}$, $\mu^{\pm}$ and $\tau^{\pm}$ respectively, and their accompanying neutrinos or anti-neutrinos. Since these particles only interact with leptons, they also fall into the category of leptophilic dark matter models~\cite{Fox:2008kb, Bertone:2008xr, John:2023ulx, John:2021ugy,Borah:2024twm, Barman:2021hhg, Nardi:2008ix, Nguyen:2025ygc}.

Within the vector dark matter context, experimental and phenomenological studies have constrained these flavor-conserving couplings as a function of dark matter mass. However, most of the constraints are in the sub-MeV regime~\cite{Shaw:2021gnp, KAGRA:2024ipf, LIGOScientific:2021ffg, Miller:2023kkd, Frerick:2023xnf, PPTA:2021uzb, Amaral:2024rbj}. For dark matter masses exceeding twice the electron mass, we provide the full branching ratios of all $B-L$ and $L_{i}-L_{j}$ particles to SM final states in Figure~\ref{fig:BR_Flavor}. We also take into account the neutrino-antineutrino final states for all decays. We assume the masses of all (anti-)neutrinos are equal and have the value $m_{\nu/\bar{\nu}}=0.01$~eV, which satisfies constraints from cosmological observations~\cite{Loureiro:2018pdz, DESI:2025ejh}. The exact neutrino masses only negligibly affect the results of our study. Notably, one can distinguish features from these vector particles compared to their dark photon counterparts, due to the fact that the neutrino decay channels are dominant in this scenario, suppressing the production of electron and positron pairs.

\section{Electrons and Positrons from Dark Matter Decay}
\label{sect:511keV}
 
Bosonic dark matter in the Milky Way can decay into SM particles. Besides neutrinos, these final states can either be $e^+e^-$ pairs, along with other charged leptons or quarks that subsequently decay into $e^+e^-$. For example, $\mu^{+}\mu^{-}$ and $\tau^{+}\tau^{-}$ pairs quickly decay to final states including $e^+e^-$ pairs on astrophysical timescales. Quark final states often hadronize to form mesons and baryons that eventually also decay to $e^+e^-$ pairs. The $e^+e^-$ source term per unit volume at the position $\Vec{r}$, taking into account all dark matter decay channels, is:

\begin{align}
\label{eq:Flux_positron}
    Q_{\rm e}(\Vec{r}, E_{\rm e})&=\sum_{f}\Gamma_{\chi\to {f\bar{f}}}\frac{\rho_{\chi}(\Vec{r})}{m_{\chi}}\left(\frac{{\rm d}N_{\rm e}}{{\rm d}E_{\rm e}}\right)\Big{|}_{\chi\to f\bar{f}}\\
    &=\Gamma_{\chi}^{\rm tot}\frac{\rho_{\chi}}{m_{\chi}}\sum_{f} {\rm BR}(\chi\to f\bar{f})\left(\frac{{\rm d}N_{\rm e}}{{\rm d}E_{\rm e}}\right)\Big{|}_{\chi \to f \bar{f}}\nonumber
\end{align}
where $\Gamma_{\chi\to {f\bar{f}}}$ is the decay rate into specific final states, while $\Gamma^\text{tot}_\chi$ is the total decay rate of the dark matter particle. These two variables are related by their decay channel branching ratios BR$\left(\chi\to f\bar{f}\right)$. 

We calculate the $e^+e^-$ spectra, ${\rm d}N_{\rm e}/{\rm d}E_{\rm e}$, for all dark matter decay channels, again calculating the total spectrum by summing over all final states weighted by their branching ratios, as

\begin{equation}
    \frac{{\rm d}N_{\rm e}}{{\rm d}E_{\rm e}}\Big{|}_{\rm tot}=\sum_{f}{\rm BR}_{\chi \to f\bar{f}}\times \frac{{\rm d}N_{\rm e}}{{\rm d}E_{\rm e}}(\chi\to f\bar{f}).
\end{equation}
\noindent 

\noindent We use the branching ratios for each dark matter model as calculated in Figs.~\ref{fig:BR_DP}, \ref{fig:BR_Flavor} and \ref{fig:BR_Scalar}. The only exception to this is the electrophilic ALP model, which decays only to $e^+e^-$ directly.

For the $e^{\pm}$ decay channel, we assume the spectrum is a Dirac distribution:
\begin{equation}
    \frac{{\rm d}N_{\rm e}}{{\rm d}E_{e}}(\chi\to {\rm e}^{+}{\rm e}^{-})=\delta\left(E_{\rm e}-\frac{m_{\chi}}{2}\right).
\end{equation}

\noindent For other decay channels, we use \texttt{Hazma}~\cite{Coogan:2019qpu, Coogan:2022cdd} and \texttt{PPPC4DMID}~\cite{Cirelli:2010xx, Ciafaloni:2010ti, Elor:2015bho, Amoroso:2018qga} to generate the $e^+e^-$  spectra. For dark matter masses below 3~GeV, we use \texttt{Hazma}\footnote{\texttt{Hazma} is named after the nuclear-type Pok\'emon
 from the fan-made Pok\'emon
 Uranium video game, which was released in 2016.} because it is the most accurate model for the $\mu^{+}\mu^{-}$ and $\pi^{+}\pi^{-}$ decay channels. We neglect the contribution from $\pi^{0}$ production, since it dominantly decays to $2\gamma$~\cite{ParticleDataGroup:2024cfk}. Above this mass range, we include flux determinations from \texttt{PPPC4DMID}, which are highly accurate for heavy dark matter (above 10~GeV). Unfortunately, the 3.5--10~GeV range is not well-probed by either simulation. Thus, to study dark matter models within this mass gap, we interpolate the results between the \texttt{Hazma} and \texttt{PPPC4DMID} for the $\mu^{\pm}$ and $\pi^{\pm}$ channels over several points within this mass range, while using \texttt{PPPC4DMID} calculations for the $\tau^+\tau^-$ decays which only occur above 2m$_\tau$. We set conservative results by taking the more conservative SM fluxes at each dark matter mass. We note that similar tools and packages that include Electroweak and QCD corrections exist~\cite{Arina:2023eic, Bauer:2020jay, Bringmann:2018lay}. However, for dark matter masses below 10~GeV, we neglect these corrections since they have negligible impact on our results.

Finally, we note that \texttt{Hazma} neglects dark matter decay to bound states that are heavier than pions, for example, $\eta$ mesons. Such final states will subsequently decay to pions and $e^+e^-$, with the pion pairs subsequently decaying to more $e^+e^-$. Ignoring these final states has a small effect on our results, but is also a conservative choice, as the $e^+e^-$ multiplicity from decays to heavy bound states always exceeds the $e^+e^-$ multiplicity from direct decays to pions. 

\subsection{Dark Matter Density in the Milky Way}

To determine the dark matter density distribution $\rho_\chi(\Vec{r})$ throughout the Milky Way, we use the generalized version of the Navarro-Frenk-White (NFW) profile~\cite{Navarro:1995iw}

\begin{equation}\label{eq: NFW profile}
\rho_\text{NFW}(r) = \rho_\text{scale}\left(\frac{r}{R_\text{scale}}\right)^{-\gamma} \left(1 + \frac{r}{R_\text{scale}}\right)^{\gamma-3},
\end{equation}
where $\rho_\text{scale}$ and $R_\text{scale}$ are the scale density and radius, $r$ is the position in the Milky Way and $\gamma$ is the profile index. We set $\rho_\odot = 0.4$~GeV/cm$^3$~\cite{Benito:2020lgu, Salucci:2010qr} and the solar position to be $R_\odot = 8.3$~kpc and set $\gamma = 1$ and $R_\text{scale} = 20$~kpc, which are typical NFW parameters~\cite{Cirelli:2010xx}.

\subsection{Probes of Low-Energy Cosmic-Ray $e^+e^-$}
\label{subsect:CRee}

After the $e^+e^-$ are injected by dark matter decay, we model their propagation through the Galaxy using the \texttt{DRAGON2} code~\cite{Evoli:2016xgn, Evoli:2017vim}. \texttt{DRAGON2} is a state-of-the-art cosmic-ray propagation code that precisely takes into account the various processes during cosmic-ray transport, such as diffusion, convection, re-acceleration and energy losses. We adopt the propagation model from Ref.~\cite{DelaTorreLuque:2023olp}. During propagation, the $e^+e^-$ thermalize due to energy losses via interactions with the interstellar medium (ISM), which are dominated by Coulomb and ionization processes for low-energy electrons. Since $e^+e^-$ are continuously produced by dark matter decay, the $e^+e^-$ flux throughout the Galaxy reaches a steady-state.

The steady state $e^+e^-$ density can either be probed directly using space-based cosmic-ray detectors, or indirectly by probing the X-rays produced via the inverse-Compton and bremsstrahlung processes. Interestingly, the premier constraint on low-mass cosmic-ray electrons stems from the Voyager-I spacecraft, whose location just outside the heliosphere makes it uniquely sensitive to low-energy cosmic rays that would be deflected by the solar wind~\cite{cummings2016galactic, stone2013voyager}. This makes Voyager-I a sensitive probe of low-energy $e^+e^-$ injections, like those expected from light bosonic dark matter~\cite{DelaTorreLuque:2023olp, DelaTorreLuque:2024qms}. A direct comparison of Voyager-I measurements, from $3$ to $\sim10$~MeV, with the $e^+e^-$ steady-state fluxes from dark matter decay sets strong limits on the lifetime of such particles, as shown in Refs.~\cite{DelaTorreLuque:2023olp, Boudaud:2016mos}.

On top of those constraints, one can use the secondary radiations emitted by the interaction of the injected $e^+e^-$ with the interstellar gas and radiation fields via bremsstrahlung and inverse-Compton scattering~\cite{DelaTorreLuque:2023huu, DelaTorreLuque:2023nhh, DelaTorreLuque:2023olp, Cirelli:2023tnx, Cirelli:2020bpc, Bartels_2017}, respectively. Therefore, observations from different X-ray telescopes, such as XMM-Newton, Nustar, Suzaku, INTEGRAL or eROSITA have been used to set strong constraints on the decay of light dark matter particles~\cite{Cirelli:2023tnx, Balaji:2025afr}. Among these, the strongest constraints are those from eROSITA and INTEGRAL (SPI). Here, we make use of our calculations of the diffuse keV X-ray background in Ref.~\cite{Balaji:2025afr} and employ the recent eROSITA diffuse data from $\sim0.3$~keV to $2.3$~keV to set our constraints, which are based on the same propagation parameters as we are using for all our constraints in this paper (for more information about the propagation parameters and uncertainties related, see Ref.~\cite{DelaTorreLuque:2023olp}). Besides the homogeneous CMB photon field, the infrared field and starlight emissions are considered here. This is computed with the \texttt{HERMES} code~\cite{Dundovic:2021ryb}, which provides the line-of-sight integrated photon emission from different directions in the sky.
To evaluate these constraints, we divide the eROSITA diffuse measurements in 30 Galactocentric, $6^{\circ}$ width, rings, as done in Ref.~\cite{Foster:2021ngm}, masking the inner $|5|^{\circ}$ around the Galactic plane. See Ref.~\cite{Balaji:2025afr} for full details.

In this work, we additionally utilize the 511~keV observations by INTEGRAL to constrain dark matter. The procedure followed here is the same as discussed in Ref.~\cite{DelaTorreLuque:2024wfz}~(see also \cite{DelaTorreLuque:2023cef, Aghaie:2025dgl, Erratum}): we compute the steady-state diffuse positron flux produced by our DRAGON models and take the positrons reaching the gas thermal energy (set to be $100$~eV). Then, the positron distribution is convolved with the cross sections for charge-exchange, which is the dominant channel for positronium production~\cite{1991ApJ...378..170G} and the Galactic electron distribution in the Galaxy, in order to determine where positronium forms. This positronium then locally annihilates to produce 511~keV line emission. Here, we focus on the longitude profile of the 511 keV emission, which is basically determined by the distribution of thermal positrons and lead to the strongest constraints for decaying sub-GeV DM~\cite{DelaTorreLuque:2023cef, Erratum}. As a last step, we compute the 511 keV flux integrated over the line-of-sight as described in Eq.~2 of Ref.~\cite{DelaTorreLuque:2023cef}.

\begin{figure}[t!]
    \centering
    \includegraphics[width=1\linewidth]{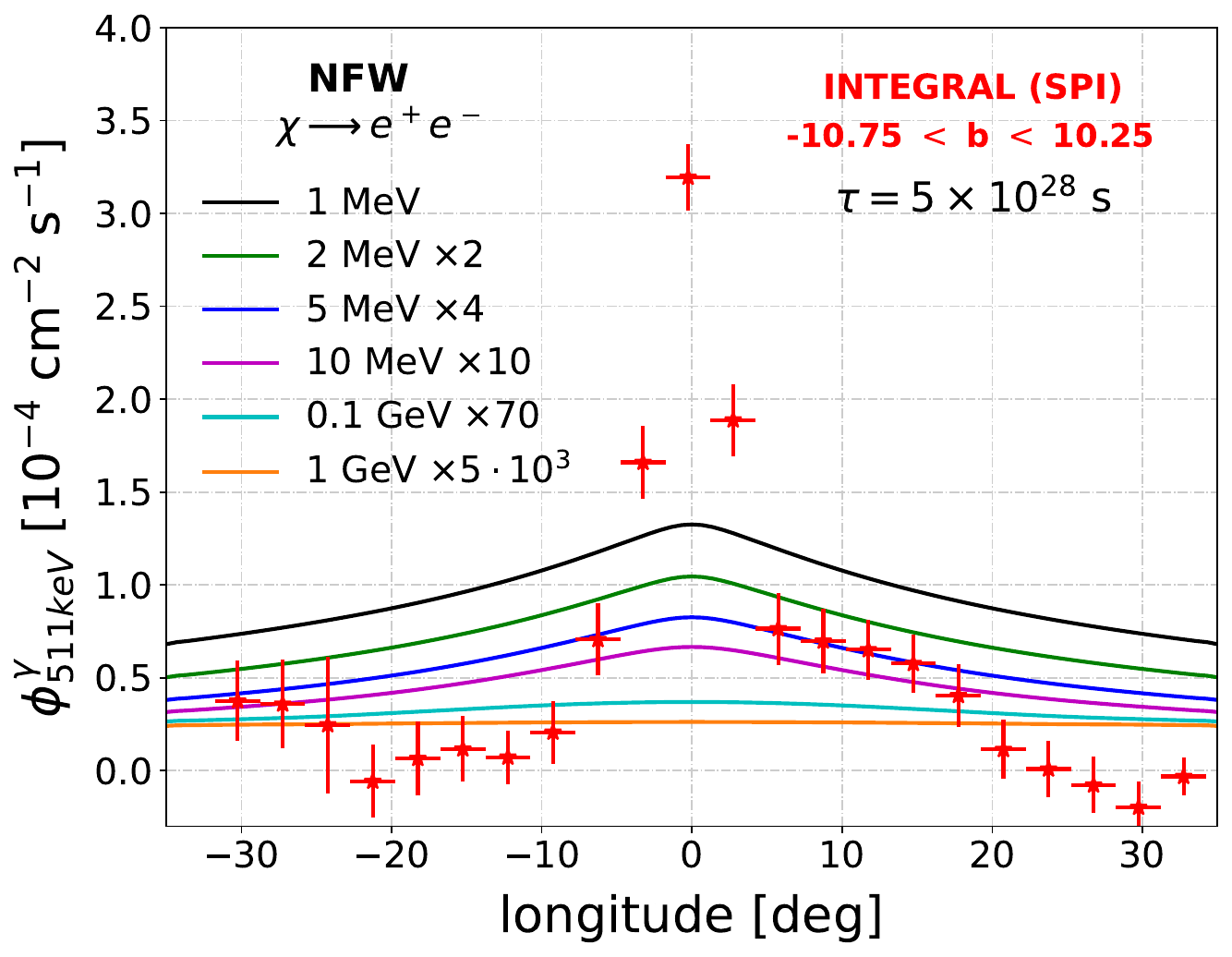}
    \vspace{-0.5cm}
    \caption{Comparison of the longitudinal profile of the $511$~keV line measured by SPI (INTEGRAL) with the expected line's profile for DM decaying with a lifetime $\tau = 5 \times10^{28}$~s, for different masses (indicated with different colors). The signals for different masses are multiplied by a constant factor that is indicated in the legend for ease of comparison. These predictions are assuming an NFW profile. } 
    \label{fig:511keV}
\end{figure}

One can directly compare the estimated 511 keV emission profile from the different DM models with SPI observations~\cite{Siegert:2015knp}. In Figure~\ref{fig:511keV} we compare the expected profiles for different DM masses with data. From this comparison, we will obtain the annihilation cross section, $\langle \sigma v \rangle$, that fits (through a $\chi^2$ fit using the { \it curve\_fit} python package) the morphology of the $511$-keV line shown in Fig.~\ref{fig:511keV}. The $95\%$ Confidence level constraints are obtained by adding the $2\sigma$ sigma error of this fit. However, we remind the reader that obtaining the morphology of the $511$-keV line is not a simple task, and requires templates to derive these data points, which add systematic uncertainties that are difficult to quantify. We note that some data points in the longitudinal profile are negative. This arises from over-subtraction errors during the template decomposition of the SPI data, as shown in Ref.~\cite{Siegert:2015knp}. These negative values are naturally accommodated in our $\chi^{2}$ fitting procedure, which produce a best-fit to the gaussian error bars of the subtracted dataset that cannot be accounted for by astrophysical foregrounds. These points are retained in the $\chi^{2}$ fit, as their uncertainties remain consistent with positive flux values, and removing them would systematically bias our results towards higher 511~keV fluxes.
To attempt to address this systematic uncertainty, we multiply the $2\sigma$ error obtained in the fit by a factor of two, which leads to a more conservative constraint. 
Moreover, we note that the uncertainties in the evaluation of these signals can be significant: varying the propagation parameters within extreme, but plausible, scenarios can lead to order-of-magnitude variations in these constraints, as shown in Ref.~\cite{DelaTorreLuque:2023cef}. We refer the reader to that work for further details, and to Fig.~\ref{fig:uncertainty} in Appendix~\ref{appen:uncertainty}, where we present the corresponding uncertainty bands for all the DM models considered here.

\begin{figure*}[t]
\centering
\includegraphics[width=2\columnwidth]{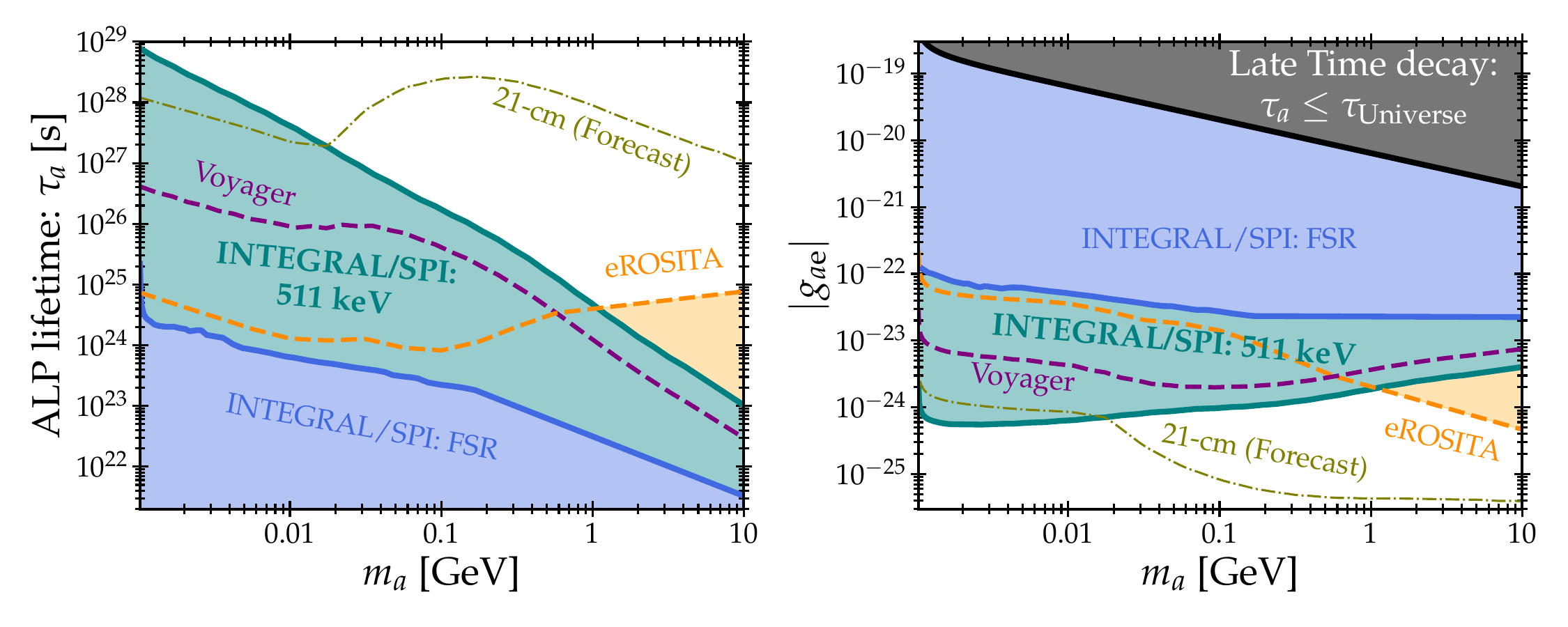}
\caption{Constraints on electrophilic ALP dark matter: ({\bf Left}) ALPs decay lifetime  and ({\bf Right}) ALPs-electron coupling $g_{a{\rm e}}$ as functions of ALP mass. Our results using 511~keV longitudinal measurement of INTEGRAL/SPI are in teal. INTEGRAL/SPI constraints using final state radiation from ALP decay from Ref.~\cite{Calore:2022pks} is in blue. Previous results consider generic dark matter decay to $e^+e^-$ are in dot-dashed lines: XMM-Newton (orange) is from Ref.~\cite{Cirelli:2023tnx}, and 21-cm HERA forecasting (olive) is from Ref.~\cite{Sun:2023acy}. Theoretical bound using the age of the Universe $\tau_{U}\simeq 4\times 10^{17}$~s is in gray.}
\label{fig:bound_ALP}
\end{figure*}

\begin{figure*}[t]
\centering
\includegraphics[width=2\columnwidth]{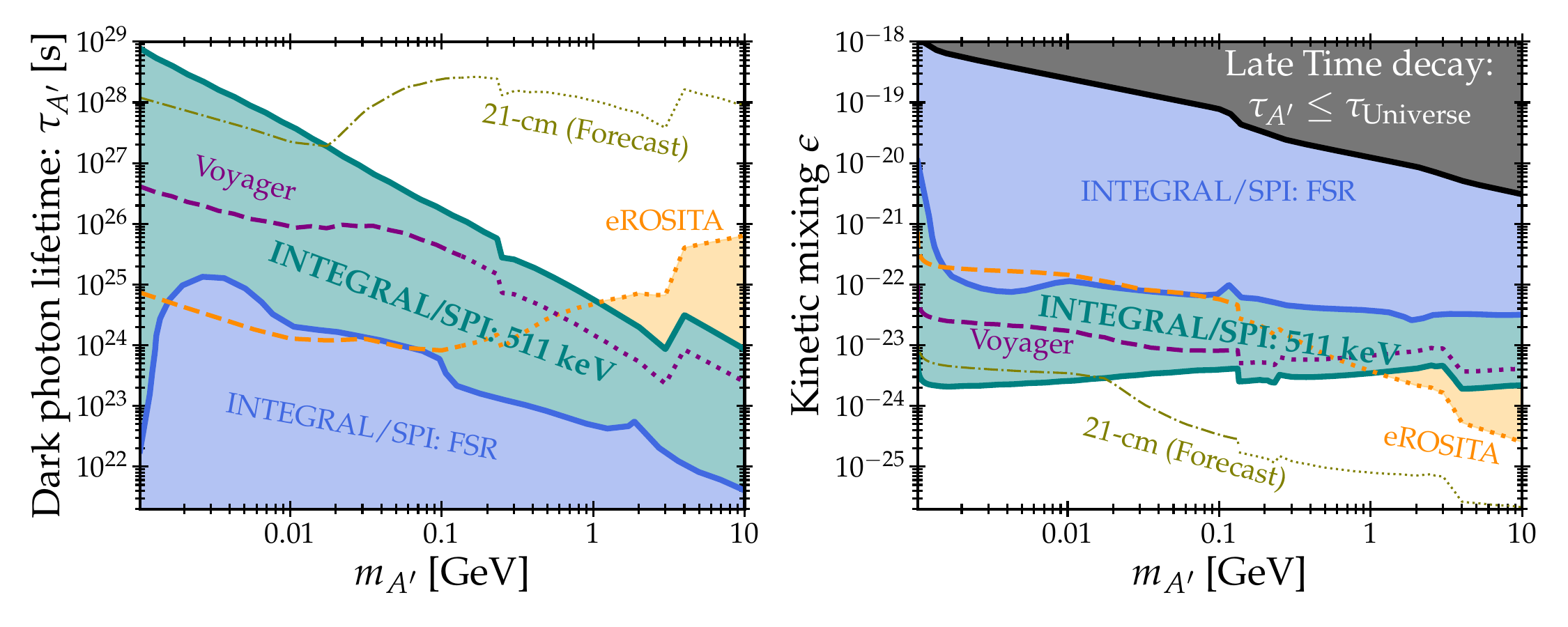}
\caption{Constraints on dark photon dark matter: ({\bf Left}) Dark photon decay lifetime  and ({\bf Right}) kinetic mixing coupling $\epsilon$ as functions of dark photon mass, similar to Figure~\ref{fig:bound_ALP}. Our estimation for dark matter above neutral pion mass scale are shown in dotted lines.}
\label{fig:bound_DP}
\end{figure*}

We calculate the diffuse electron/positron spectra and secondary emissions for these DM particles in the 1~MeV to 10~GeV mass range. Focusing on the longitude profile of the $511$-keV line, we derive 95\%~C.L. lower limits of the (total) decay lifetime in this mass interval for each particle model, accounting for the different electron injections expected above twice the muon mass.   
Meanwhile, for the constraints from Voyager-1 and eROSITA data we make use of the previously derived constraints~\cite{DelaTorreLuque:2023olp, Balaji:2025afr} on direct decay into $e^+e^-$, and scale the limits above twice the muon mass to account for the different $e^{+}e^{-}$ injection yields. We apply the same strategy to calculate the projected limit from 21-cm line measurement from the future HERA experiment~\cite{DeBoer_2017} that considers direct decay to $e^+e^-$ final state~\cite{Sun:2023acy}. We notice that our estimation above twice the muon mass threshold may slightly different with the actual analysis using the projected HERA sensitivity of the 21-cm spectrum~\cite{DeBoer:2016tnn}. From these lower limits on the lifetime, we compute the upper limits of the characteristic couplings for these DM particles as functions of the DM mass.

\section{Results}
\label{sect:result}

\begin{figure*}[t]
\centering
\includegraphics[width=2\columnwidth]{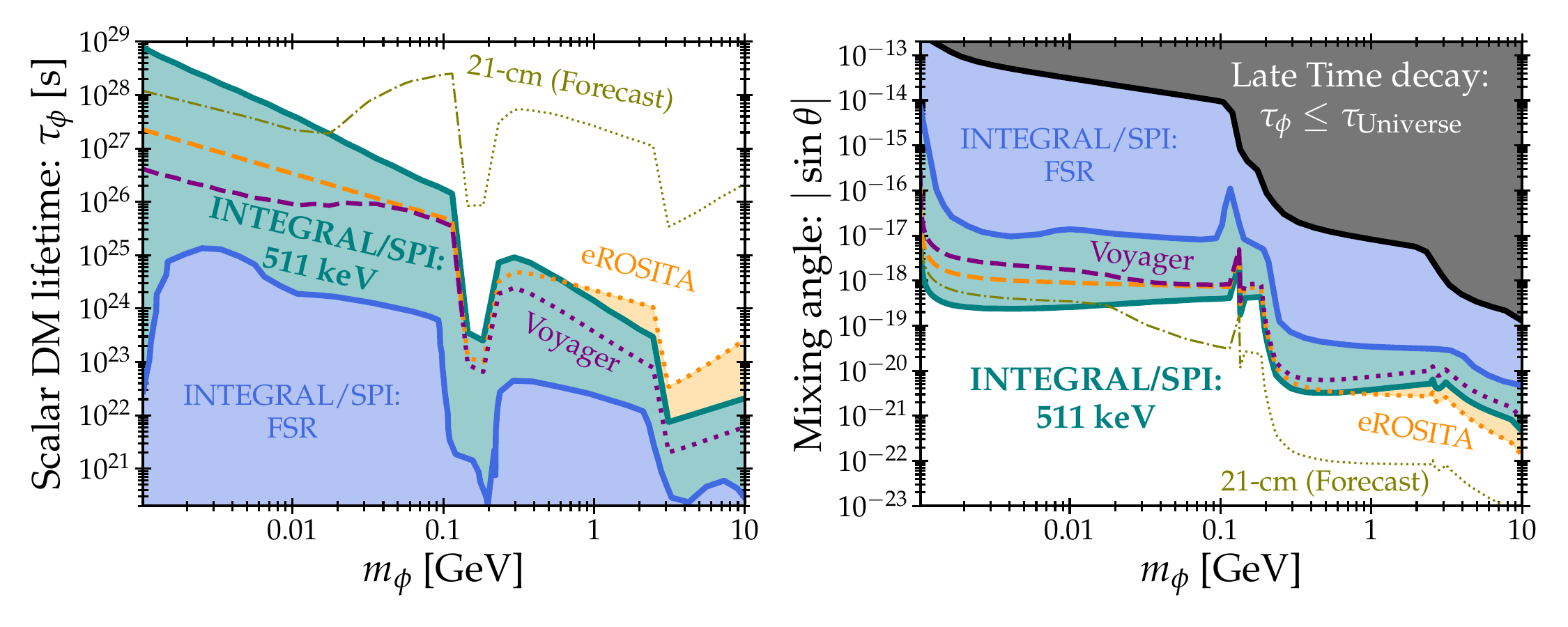}
\caption{Constraints on scalar dark matter: ({\bf Left}) Scalar dark matter decay lifetime  and ({\bf Right}) mixing angle as functions of the scalar dark matter mass, similar to Figs.~\ref{fig:bound_ALP} and \ref{fig:bound_DP}.}
\label{fig:bound_Scalar}
\end{figure*}

\begin{figure*}[t]
\centering
\includegraphics[width=2\columnwidth]{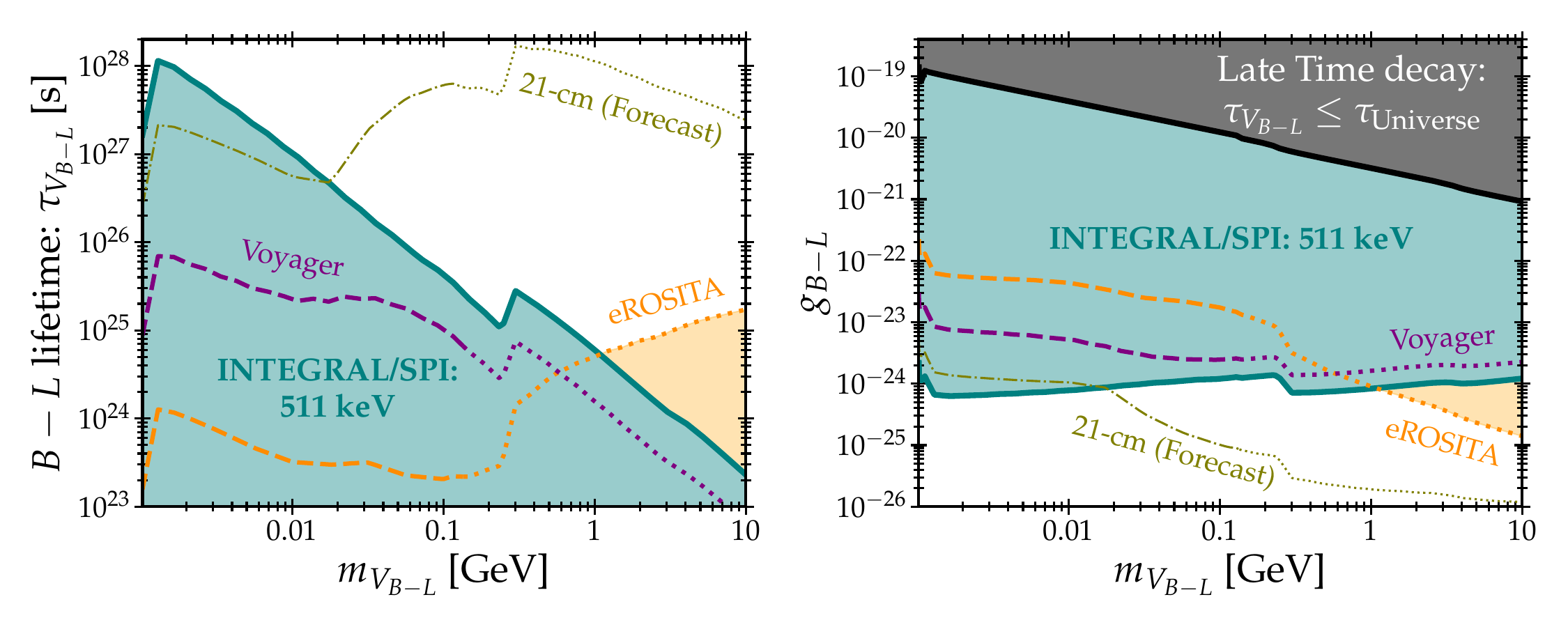}
\caption{Constraints on the universal $B-L$ vector dark matter: ({\bf Left}) $B-L$ decay lifetime and ({\bf Right}) $g_{B-L}$ coupling as functions of vector dark matter mass. The gray theoretical bound consider dark matter has a lifetime more than the universe age considering the dominant $V\to \nu\bar{\nu}$ decay channel from Ref.~\cite{Chun:2022qcg}.}
\label{fig:bound_BL}
\end{figure*}

\begin{figure*}[t]
\centering
\includegraphics[width=2\columnwidth]{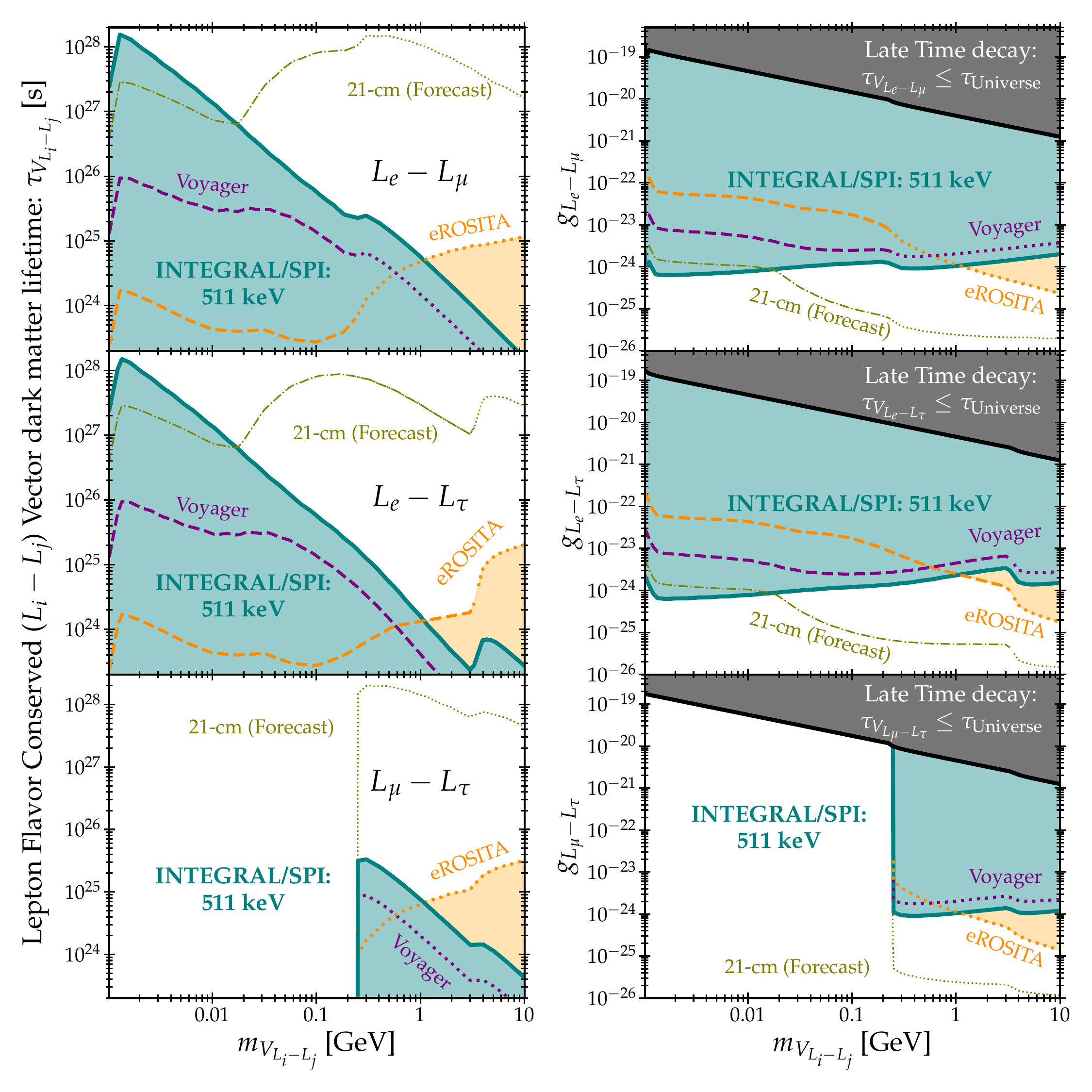}
\caption{Constraints on lepton-flavor-dependent $L_{i}-L_{j}$ dark matter model ({\bf Left}) decay lifetime and ({\bf Right}) coupling as functions of vector dark matter mass. Top, mid and bottom panels are for constraints on $L_{e}-L_{\mu}$, $L_{e}-L_{\tau}$, and $L_{\mu}-L_{\tau}$ respectively. We calculate the theoretical bounds in gray similar to Fig~\ref{fig:bound_BL}, considering $V\to\nu\bar{\nu}$ decay channels.}
\label{fig:bound_LL}
\end{figure*}

We present our constraints on the dark matter decay lifetimes (left panels) and the corresponding coupling constants (right panels) as functions of the dark matter mass in Figures~\ref{fig:bound_ALP}, \ref{fig:bound_DP}, \ref{fig:bound_Scalar}, \ref{fig:bound_BL} and \ref{fig:bound_LL} for each of our theoretically motivated dark matter models. In each figure, the results from the INTEGRAL analysis of the 511 keV line are shown in teal, constraints from the eROSITA analysis of inverse-Compton scattering are shown in orange, and constraints from the Voyager cosmic-ray $e^+e^-$ flux are shown in purple. We compare our results with theoretical constraints requiring the dark matter particles to have lifetimes exceeding the age of the universe ($\sim$13.8 Gyr) (gray shaded regions), previous INTEGRAL constraints on final-state radiation (FSR) from dark matter decay (royal blue shaded regions)~\cite{Calore:2022pks, Nguyen:2024kwy}, as well as forecasts for future constraints from 21~cm observations~\cite{Sun:2023acy} (olive). 

We note that for dark matter masses below the pion mass, only the $e^+e^-$ final states are relevant. At larger masses, bosonic dark matter candidates (excluding electrophilic ALPs) can also decay into other Standard Model particles. We estimate the corresponding constraints (for eROSITA and Voyager) and projections (21-cm) by computing the fraction of final-state $e^+e^-$ produced both directly and via secondary decays from other SM channels. These fractions are then used to rescale the original bounds. For INTEGRAL data, we calculate the exact bounds by directly calculating the population of thermal positrons. Throughout these figures, solid lines indicate that the constraints are calculated exactly for the SM final states under investigation, while dotted lines indicate that the results are interpolated by combining previously calculated constraints that consider single $e^+e^-$ final states. Dashed lines (constraints) and dot-dashed lines (forecast) indicate regions where only the $e^+e^-$ final state is relevant, making this interpolation exact.

Figures~\ref{fig:bound_ALP}, \ref{fig:bound_DP}, and \ref{fig:bound_Scalar} present our constraints on the decay of axion-like particles, dark photons, and scalar dark matter. For ALPs, we constrain lifetimes in the range $10^{23}$--$10^{29}$~s and axion-electron couplings down to $10^{-24}$ for masses between 1~MeV and 10~GeV. Our limits, derived from the 511 keV line observed by INTEGRAL/SPI, improve upon previous results that utilized INTEGRAL continuum data~\cite{Calore:2022pks} by 3--4 orders of magnitude in lifetime and roughly 2 orders of magnitude in coupling strength. At lower masses, the improvement stemming from utilizing 511~keV line data is even more pronounced, as the the intensity of the line signal remains similar, while the FSR contribution to the continuum fades. In addition to this INTEGRAL constraint, ALPs above 1~MeV primarily decay to $e^+e^-$ (with the $\gamma\gamma$ channel loop-suppressed), we directly compare our results to dark matter decay into $e^+e^-$ final states from eROSITA\cite{Balaji:2025afr} and Voyager~\cite{DelaTorreLuque:2023olp} along with projected 21-cm constraints~\cite{Sun:2023acy}. We find that while Voyager constraints are always subdominant to 511~keV line constraints, eROSITA data can produce stronger constraints for dark matter masses above 1~GeV, where the continuum emission becomes brightest.

For dark photons (Figure~\ref{fig:bound_DP}), we constrain lifetimes from $10^{24}$–$10^{29}$~s and kinetic mixing parameters $\epsilon$ down to $10^{-24}$, improving upon previous INTEGRAL/SPI continuum constraints from Ref.~\cite{Nguyen:2024kwy} by 2--4 orders of magnitude in lifetime and 1--2 orders in $\epsilon$. In this scenario, the eROSITA and Voyager constraints can be directly compared to previous $e^+e^-$ constraints below 1.1~MeV, and are extrapolated at higher dark matter masses. We find again that Voyager constraints remain subdominant to 511~keV line constraints, while eROSITA constraints can be dominant for larger dark matter masses.

Finally, our scalar dark matter constraints (Figure~\ref{fig:bound_Scalar} extend to lifetimes of $10^{22}$--$10^{29}$~s and mixing angles $\sin\theta \simeq \theta$ down to $10^{-19}$–$10^{-21}$, also improving prior bounds by 2--4 orders in lifetime and 1--2 orders in mixing angle. In scalar models, we note a significant weakening in the dark matter constraint at the neutral pion mass ($\sim$135~MeV) because scalar dark matter in this mass range dominantly decays to neutral pions that do not produce $e^+e^-$ final states. At higher masses (above twice the muon mass), final states that produce $e^+e^-$ again become prevalent, strengthening the constraints. As in previous results, we find that Voyager constraints are subdominant, while eROSITA dominates constraints above a few GeV.

Moreover, we show the constraints on flavor-universal $B-L$ vector dark matter in Fig.~\ref{fig:bound_BL}, and constraints on the three lepton-flavor-dependent vector dark matter particles in Fig.~\ref{fig:bound_LL}. Since previous analysis of INTEGRAL FSR data did not consider these vector dark matter particles, there are no projected constraints from those works to display. These particles can decay to neutrino-antineutrino pairs with branching ratios that can go up 80--85~\%, which generally makes the constraints on these dark matter models weaker by a factor of $\sim$5 compared to other classes of dark matter models.
Therefore, our constraints for $B-L$, $L_{e}-L_{\mu}$, and $L_{e}-L_{\tau}$ particles on their lifetime are from $10^{23}$--$10^{28}$~s, with corresponding couplings $g_{B-L}$, $g_{L_{e}-L_{\mu}}$, and $g_{L_{e}-L_{\tau}}$ that extend down to $10^{-24}$. Notably, for the $L_{\mu}-L_{\tau}$ vector dark matter, we only have the constraints above twice the mass of muon particle due to kinematic considerations, with the limit on the dark matter lifetime that spans from $10^{23}$--$10^{26}$~s, and couplings $g_{L_{\mu}-L_{\tau}}$ that extend down to $10^{-24}$. We additionally show Voyager and eROSITA constraints, the latter of which can exceed 511 keV dark matter constraints above $\sim$1~GeV.

Finally, in all cases we compare our results with projections for HERA 21~cm observations~\cite{Sun:2023acy}, finding that these future datasets can significantly constrain light bosonic dark matter. One consideration regarding the robustness of 21~cm line projected constraints is the uncertainty in the astrophysical models involved, including: the star formation efficiency, their UV/X-ray photon emission, and the spectrum of Lyman-$\alpha$ photons. The potential of such signals thus motivates further efforts to explore these astrophysical uncertainties further~\cite{Cima:2025zmc}. These projections also strongly depend on our models for the first galaxies, providing the exciting possibility that JWST observations, which challenge models for the first galaxies~\cite{nakane2025feabundancesearlygalaxies, Maiolino_2024, adams2024epochspaperiiultraviolet}, further improve the robustness of 21~cm constraints. Finally, in the case of very light dark matter particles (near twice the electron mass), 511~keV constraints are predicted to exceed even next-generation 21~cm instrumentation. The complementarity between these constraints motivates future explorations of next-generation 511~keV detectors, such as COSI~\cite{Tomsick:2023aue}.

\section{Discussion}
\label{sect:discussion}

In this section, we note several important trends in the strength of various probes of bosonic dark matter decay. We note that 511~keV line constraints are generally strongest for dark matter particles which lie just above twice the electron mass. This is due to the fact that the barely relativistic positrons thermalize much more efficiently in the interstellar medium, enhancing the probability of positronium formation and the strength of the resulting line signal. Moreover, additional decays to non-electron channels (which then decay to $e^+e^-$ pairs), tend to further enhance the strength of 511~keV line constraints, as these secondary decays induce a higher multiplicity of cooler electrons compared to direct $e^+e^-$ decays.

The energy scaling of Voyager constraints is similar. The strongest constraints stem from a high multiplicity of $e^+e^-$ pairs, rather than from a high energy-flux of such final states -- though we note some deviations from this model based on the energy-dependent effective area of Voyager instrumentation. While Voyager constraints are typically weaker than 511~keV line constraints, they are complementary as they are based on an entirely different set of propagation assumptions and astrophysical background models. We note that at higher energies (above 10~GeV), $e^+e^-$ pairs begin to efficiently traverse the heliosphere and arrive at Earth-orbiting instruments such as AMS-02~\cite{amspaper}. The much larger collecting area (and better energy reconstruction) of these instruments produce much stronger constraints above these mass ranges~\cite{Nguyen:2024kwy}. We note that AMS-02 could potentially continue to provide world-leading constraints all the way down to dark matter masses near 1~GeV, though this would require significantly improved models of solar modulation~\cite{Kuhlen:2019hqb}.

eROSITA constraints, on the other hand, typically depend on the total amount of energy injected through bremsstrahlung, inverse-Compton scattering and other sub-dominant processes. Thus, in this sub-GeV regime, the constraints weaken less significantly when the dark matter particle becomes heavier. This indicates that we should generally expect eROSITA to perform best at heavy dark matter masses, as we see in our results. The same is true for the projected 21~cm constraints, where the total energy injected by dark matter decays (typically produced via inverse-Compton scattering) is relevant, rather than the total number of decays. We note that our limits in both scenarios are conservative, as we consider only energy-losses from $e^+e^-$ pairs. In reality, additional radiation is produced via the energy-losses of muons, as well as neutral and charged pions. We expect such processes to strengthen our limits by less than a factor of two for most dark matter masses and final states, but leave the complete calculation of these effects to future work.

\section{Conclusion and outlook}
\label{sect:conclusion}

In this paper, we present a detailed analysis of the constraints on four theoretically well-motivated models of bosonic dark matter: electrophilic axion-like particles, dark photons, scalar particles and vector bosons with either flavor-universal or lepton-flavor-dependent couplings. For each model, we compute the complete set of branching ratios into all accessible Standard Model final states, taking into account the full kinematic and interaction structure of the decays. We then generate the resulting electron and positron spectra, incorporating contributions from both direct decays and secondary processes. To model the propagation of these $e^+e^-$ particles through the Galaxy, we employ the \texttt{DRAGON} code, carefully accounting for diffusion, energy losses, radiative emission and positronium formation, which gives rise to the distinctive 511 keV $\gamma$-ray line. 

We constrain these models using a combination of INTEGRAL observations of the 511~keV line, Voyager observations of the low-energy $e^+e^-$ flux, and eROSITA constraints on the X-ray continuum emission. We also examine projected sensitivities for future 21~cm observations with HERA. 

Our constraints on bosonic dark matter models cover a mass range from 1~MeV to 10~GeV, with lower bounds on the dark matter lifetime ranging from $10^{23}$--$10^{29}$~s. For axion-like particles, dark photons, and lepton-flavor-dependent vector dark matter, we constrain the relevant couplings to SM particles down to $10^{-24}$. In the case of scalar dark matter, we derive limits on the mixing angle, reaching values as low as $10^{-19}$–$10^{-21}$. These results stand as world-leading limits on bosonic dark matter decay, often exceeding previous constraints by 2--5 orders of magnitude. Additionally, our results highlight the importance of accounting for the full decay spectrum, including all electron/positron and photon channels, in future analyses. A more rigorous treatment of these channels could substantially strengthen constraints above the neutral pion mass.

\begin{figure*}[tbp!]
\centering
\includegraphics[width=2\columnwidth]{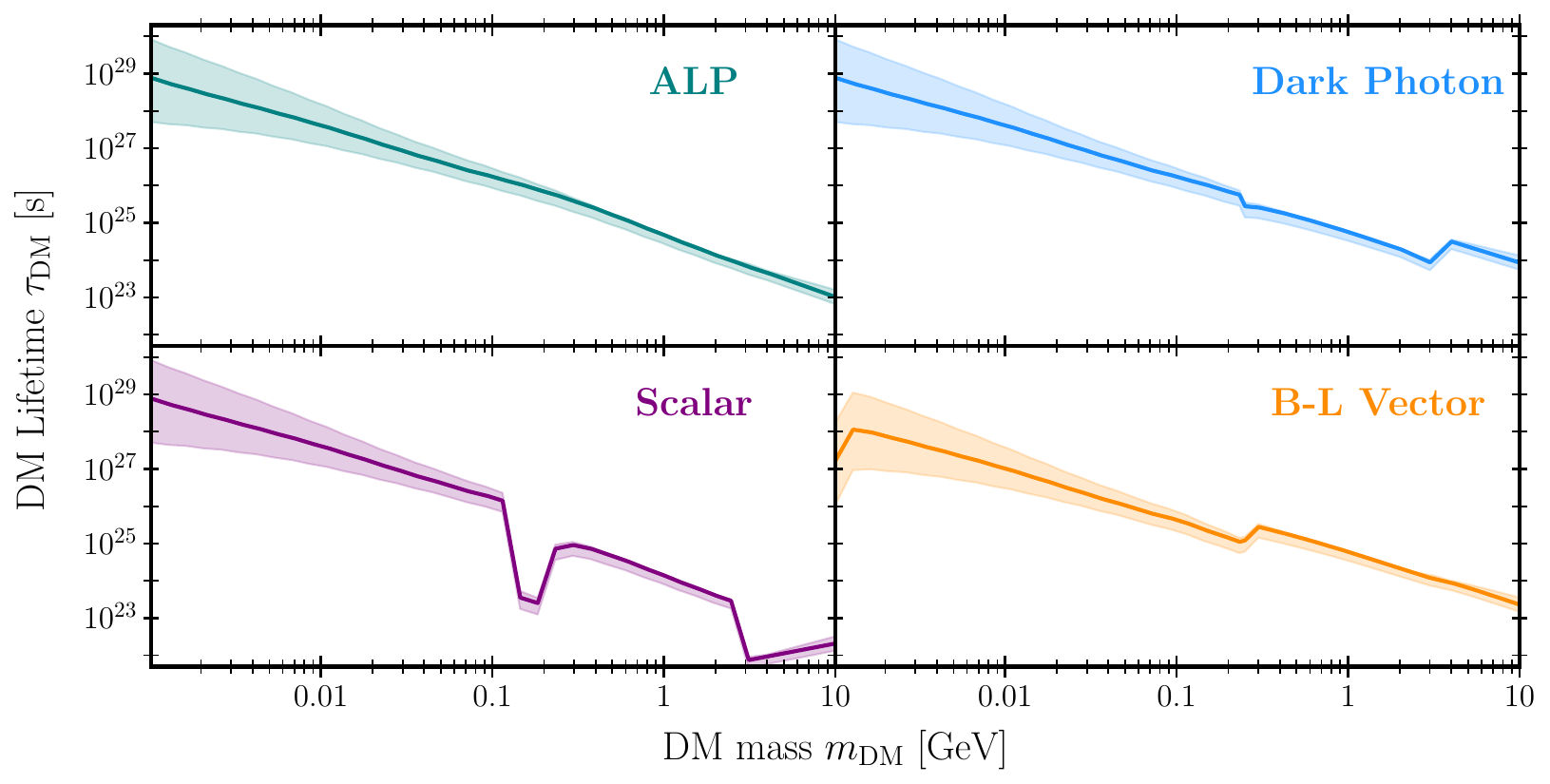}
\caption{Propagation uncertainties on the INTEGRAL 511-keV constraints on the DM decay lifetime for four bosonic DM models: electrophilic ALP (top left, teal), dark photon (top right, blue), scalar DM (bottom left, purple), and $B-L$ vector DM (bottom right, orange). Solid lines show the 95\% CL lower limits, and shaded bands represent the uncertainty from varying the propagation parameters.}
\label{fig:uncertainty}
\end{figure*}

We note that our constraints on the dark matter lifetimes decrease from $10^{29}$~s to $10^{23}$~s as the dark matter mass increases from 1~MeV to 10~GeV across all considered bosonic dark matter models. The weakening of these constraints is a feature of the well-known ``MeV-gap" that stems from the lack of sensitive instruments in the MeV range. Future MeV-range gamma-ray telescopes~\cite{Cirelli:2025qxx, O_Donnell_2025}, such as VLAST~\cite{tang2025prospectsprobingsubgevleptophilic, VLASTpaper}, MeVCube~\cite{Lucchetta:2022nrm, Lucchetta:2022zrd, Saha:2025wgg}, AMEGO-X~\cite{AMEGO:2019gny}, e-ASTROGAM~\cite{e-ASTROGAM:2017pxr}, GECCO~\cite{Orlando:2021get}, MAST~\cite{Dzhatdoev:2019kay}, GRAMS~\cite{Aramaki:2019bpi}, and PANGU~\cite{Wu:2014tya} would provide significantly improved limits on these classes of dark matter models. In particular, the upcoming COSI mission, scheduled for launch in 2027, will provide high-precision observations of the 511~keV line, offering the potential for significantly improving constraints on these dark matter scenarios~\cite{Watanabe:2025pvc}.

Finally, our analysis framework can be extended to probe dark matter annihilation scenarios. Many dark matter models include portal particles, such as axion-like particles, dark photons, scalars with Yukawa couplings, or vector bosons associated with $B-L$ and $L_i-L_j$ symmetries that mediate interactions between the dark sector and the Standard Model. These portals can produce electron/positron spectra with similar shapes and branching ratios to those considered in this work~\cite{DEramo:2025jsb, CirelliPrep}. As such, our pipeline based on the 511~keV line can impose stringent constraints on the couplings and masses of these portal particles, offering insights into the possible production mechanisms of the associated dark matter candidates.

\section*{Acknowledgment}
TTQN thanks Maurico Bustamante, Markus Ahlers and the Neils Bohr Institute for their hospitality during the NBIA PhD school  ``Neutrinos: Here, There, Everywhere". We thank Tim Tait for helpful discussions regarding dark photon and vector dark matter model, Tim Tait and Kevin Zhou for discussions regarding scalar dark matter, Seokhoon Yun on $B-L$ dark matter bound, Arianne Dekker for cross checking our ALPs constraints, as well as Arthur Loureiro and Vivian Poulin for the discussions regarding the neutrino mass. We also thank Joshua Foster and Tracy Slatyer for helpful comments regarding 21~cm forecasts.  SB is supported by the STFC under grant ST/X000753/1. PDL is supported by the Juan de la Cierva JDC2022-048916-I grant, funded by MCIU/AEI/10.13039/501100011033 European Union "NextGenerationEU"/PRTR. The work of PDL is also supported by the grants PID2021-125331NB-I00 and CEX2020-001007-S, both funded by MCIN/AEI/10.13039/501100011033 and by ``ERDF A way of making Europe''. PDL also acknowledges the MultiDark Network, ref. RED2022-134411-T. 
TTQN, PC and TL are supported by the Swedish Research Council under contract 2022-04283. TL also acknowledges support from the Wenner-Gren foundation under grant SSh2024-0037. IJ acknowledges support from the Research grant TAsP (Theoretical Astroparticle Physics) funded by INFN, and Research grant ``Addressing systematic uncertainties in searches for dark matter'', Grant No.\ 2022F2843L, CUP D53D23002580006 funded by the Italian Ministry of University and Research (\textsc{mur}).

This work made use of {\tt Numpy}~\cite{Harris_2020}, {\tt SciPy}~\cite{Virtanen:2019joe}, {\tt astropy}~\cite{Astropy:2013muo}, {\tt matplotlib}~\cite{HunterMatplotlib}, {\tt Jupyter}~\cite{2016ppap.book...87K}, as well as {\tt Webplotdigitizer}~\cite{Rohatgi2022}.

\appendix

\section{Propagation Uncertainties on the 511-keV Constraints}
\label{appen:uncertainty}

The constraints derived from the INTEGRAL 511-keV line depend on the modeling of cosmic-ray propagation through the Galaxy. In particular, the diffusion coefficient, the size of the diffusion zone, convection, and re-acceleration parameters all affect the steady-state positron distribution. Consequently, they also affect the rate of positronium formation and 511-keV emission. As discussed in Ref.~\cite{DelaTorreLuque:2023cef}, varying these propagation parameters within extreme, but plausible, scenarios can lead to order-of-magnitude uncertainties in the resulting constraints.

In Fig.~\ref{fig:uncertainty}, we show the effect of propagation uncertainties on the 511-keV lifetime constraints for four representative DM models: electrophilic ALPs, dark photon, scalar DM, and $B-L$ vector DM. The shaded bands are obtained by multiplying the benchmark lifetime limits by the upper and lower uncertainty factor, which is calculated following the range of propagation models considered in Ref.~\cite{DelaTorreLuque:2023cef}. Above 5~GeV, the uncertainty factors are approximately unity. However, for lower DM masses, the variation in the limit can exceed an order of magnitude in either direction, making the constraint dependent on the specific  propagation model that we consider. 

We note that similar propagation uncertainties also affect the Voyager and eROSITA constraints presented in the main text. A detailed discussion of those uncertainties can be found in Refs.~\cite{DelaTorreLuque:2023olp, Balaji:2025afr}.

\bibliography{ref}

\end{document}